\documentclass[11pt,twoside,a4paper,reqno]{amsart}

\usepackage{amssymb,amscd,upref,mathrsfs,url}
\usepackage{hyperref}
\let\mathcal\mathscr
\let\mathcal\mathscr
\let\kappa\varkappa
\let\phi\varphi

\DeclareFontFamily{OML}{cyi}{} \DeclareFontShape{OML}{cyi}{m}{n}{
  <5> <6> <7> <8> <9> gen * wncyi
  <10> <10.95> <12> <14.4> <17.28> <20.74> <24.88> wncyi10
 }{}
\DeclareSymbolFont{rusletters}{OML}{cyi}{m}{n}
\DeclareSymbolFontAlphabet{\rusmath}{rusletters}
\DeclareMathSymbol\re{\rusmath}{rusletters}{"03}

\newcommand{\bom}{{\bf \Omega}}
\newcommand{\bop}{{\bf \Pi}}
\newcommand{\cprime}{\/{\mathsurround=0pt$'$}}
\newcommand*{\pd}[2]{\frac{\partial#1}{\partial#2}}
\newcommand*{\abs}[1]{\lvert#1\rvert}
\renewcommand*{\d}{\mathinner{\!}d}
\def\cprime{\/{\mathsurround=0pt$'$}}
\DeclareMathOperator{\sym}{sym}
\DeclareMathOperator{\Cl}{Cl}
\DeclareMathOperator{\cosym}{cosym}
\DeclareMathOperator{\Hom}{Hom}
\newcommand*{\eval}[1]{\left.#1\right|}
\newcommand*{\sd}[2]{\{\,#1\mid#2\,\}}

\newcommand{\ldb}{[\![}
\newcommand{\rdb}{]\!]}

\allowdisplaybreaks[4]

\theoremstyle{remark}
\newtheorem{remark}{Remark}

\begin{document}
\title[Integrable structures for a Monge-Amp\`{e}re equation]{On integrable
  structures for a generalized Monge-Amp\`{e}re equation}

\thanks{This work was supported in part by the NWO-RFBR grant 047.017.015 (PK,
  IK \& AV), RFBR-Consortium E.I.N.S.T.E.IN grant 09-01-92438 (IK, AV \& RV)
  and RFBR-CNRS grant 08-07-92496 (IK \& AV)}

\author{P. Kersten} \address{Paul Kersten \newline
  University of Twente, Faculty of Mathematical Sciences \\ P.O.~Box 217 \\
  7500 AE Enschede \\ The Netherlands (on retirement since 2010)}
\email{kersten@math.utwente.nl}

\author{I. Krasil{\cprime}shchik}
\address{Iosif Krasil{\cprime}shchik \newline
  Independent University of Moscow \\
  B. Vlasevsky~11 \\ 119002 Moscow \\ Russia} \email{josephk@diffiety.ac.ru}

\author{A.~Verbovetsky}
\address{Alexander Verbovetsky \newline Independent University of Moscow \\
  B. Vlasevsky~11 \\ 119002 Moscow \\ Russia}

\email{verbovet@mccme.ru}

\author{R.~Vitolo} \address{Raffaele Vitolo\newline Dept.\ of Mathematics
  ``E. De Giorgi'', Universit\`a del Salento, via per Arnesano, 73100 Lecce,
  Italy} \email{raffaele.vitolo@unisalento.it}


%\subjclass[2000]{37K05, 35Q53}

\begin{abstract}
  We consider a $3$rd-order generalized Monge-Amp\`{e}re
  equation~$u_{yyy}-u_{xxy}^2 +u_{xxx} u_{xyy}=0$ (which is closely related to
  the associativity equation in the $2$-d topological field theory) and
  describe all integrable structures related to it (i.e., Hamiltonian,
  symplectic, and recursion operators). Infinite hierarchies of symmetries and
  conservation laws are constructed as well.
\end{abstract}
\maketitle

\section*{Introduction}
\label{sec:introduction}

Monge-Amp\`{e}re equations~\cite{KuLyRu-CUP-2007} is one of the most
interesting objects to apply methods of geometrical theory of differential
equation. Generalizations of classical Monge-Amp\`{e}re equations are
discussed, e.g., in~\cite{Bo-CRAS-1992}. One of such generalizations is the
equation
\begin{equation}\label{eq:23}
  u_{yyy}-u_{xxy}^2 +u_{xxx} u_{xyy}=0.
\end{equation}
This is a third-order Monge-Amp\`{e}re equation
(\cite{Bo-CRAS-1992,KuLyRu-CUP-2007}), but this does not help too much in
understanding its integrability properties.

Equation~\eqref{eq:23} is closely related to the associativity equation in
$2$-d topological field theory~\cite{Du-LNM-1996} and was studied in a number
of papers
(\cite{FeGaMoNu-CMPh-1997,FeMo-FAA-1996,KaNu-PhLetA-1997,KaNu-JPhA-1998,St-PhLetA-1996})
and its integrability (existence of a bi-Hamiltonian structure) was
established.

Note though that in these papers the equation was not considered in the
initial form~\eqref{eq:23}, but was rewritten as a three-component system
\begin{equation}
  \label{eq:24}
  a_y=b_x,\quad b_y=c_x,\quad c_y=(b^2-ac)_x
\end{equation}
of hydrodynamical type. Of course, equations~\eqref{eq:23} and~\eqref{eq:24}
are closely related, but not the same and even not equivalent being associated
to each other by the differential substitution
\begin{equation*}
  a=u_{xxx},\qquad b=u_{xxy},\qquad c=u_{xyy}
\end{equation*}
(just like the KdV and mKdV are related by the Miura map or the Burgers and
heat equations by the Cole-Hopf transformation).

The aim of this paper is to attack Equation~\eqref{eq:23} directly, not
reducing it to the evolutionary form, and to study the structures that arise
on this equation. To this end, we use geometrical and cohomological methods
described initially in~\cite{KeKrVe-JGP-2004} and discussed in detail in a
recent review paper~\cite{KrVe-JGP-2011}. These methods has been successfully
applied to a number of equation (see,
e.g.,~\cite{GoKrVe-AAM-2008,KeKrVe-AAM-2006}).

In Section~\ref{sec:theor-backgr} we briefly recall basic notions from the
geometry of jet spaces. Section~\ref{sec:main-results} contains main results
on the Monge-Amp\`{e}re equation~\eqref{eq:23} (including description of
Hamiltonian, symplectic, and recursion operators, as well hierarchies of
symmetries and conservation laws). In particular, we show that
Equation~\eqref{eq:23} admits a symplectic structure of the form~$D_x$ (this
is the only \emph{local} operator that is responsible for the integrability of
the equation and it corresponds to the symplectic structure described
in~\cite{Witten-1989,Zucker-1986}). A nonlocal Hamiltonian
structure~$D_x^{-1}$ corresponds to this operator. Other operators are quite
complicated and are described in
Sections~\ref{sec:ch-appl:--rec-sym-2-ass},~\ref{sec:ch-appl:-symp-2-ass},~\ref{sec:ch-appl:-Ham-2-ass},
and~\ref{sec:ch-appl:--rec-cosym-2-ass}.

All computations were done using~\texttt{CDIFF}, a \texttt{REDUCE} package for
computations in geometry of differential equations
(see~\url{http://gdeq.org}).

\section{Theoretical background}
\label{sec:theor-backgr}

\subsection{Jets and equations}
\label{sec:jets-equations}

Recall that geometric approach to PDEs~\cite{BoChDu-AMS-1999} assumes that an
equation~$\mathcal{E}$ together with all its prolongations (i.e., differential
consequences) is a submanifold in the manifold~$J^\infty(\pi)$ of infinite
jets of some bundle~$\pi\colon E\to M$, where~$M$ and~$E$ are smooth manifolds
of dimensions~$n$ and~$n+m$, respectively.

The first manifold is the one that contains independent variables, while the
sections of~$\pi$ play the role of unknown functions (fields)
in~$\mathcal{E}$. If~$U\subset M$ is a coordinate neighborhood such
that~$\eval{\pi}_U$ is trivial then we choose local
coordinates~$x^1,\dots,x^n$ in~$U$ and~$u^1,\dots,u^m$ in the fiber
of~$\eval{\pi}_U$. Then the corresponding \emph{adapted
  coordinates}~$u_\sigma^j$, $\sigma$ being a multi-index, in~$J^\infty(\pi)$
are defined as follows. For a local section~$f=(f^1,\dots,f^m)$ we set
\begin{equation*}
  f^*(u_\sigma^j)=\pd{^{\abs{\sigma}}f^j}{x^\sigma}.
\end{equation*}
\emph{Functions} on~$J^\infty(\pi)$ may depend on~$x^i$ and \emph{finite}
number of~$u_\sigma^j$ only.

The vector fields
\begin{equation*}
  D_i=\pd{}{x^i}+\sum_{j,\sigma}u_{\sigma i}^j\pd{}{u_\sigma^j}\qquad i=1,\dots,n
\end{equation*}
are called \emph{total derivatives} and differential operators in total
derivatives are called \emph{$\mathcal{C}$-differential operators}.

If an equation is given by the system~$F=0$, where~$F=(F^1,\dots,F^r)$ is a
vector-function on~$J^\infty(\pi)$, then its \emph{infinite
  prolongation}~$\mathcal{E}$ is given by
\begin{equation*}
  D_\sigma(F)=0,\qquad \abs{\sigma}\ge 0,
\end{equation*}
where~$D_\sigma=D_{\sigma_1}\circ\dots\circ D_{\sigma_s}$
for~$\sigma=\sigma_1\dots\sigma_s$. Total derivatives can be restricted
to~$\mathcal{E}$ (we preserve the same notation for these restrictions) and
generate the \emph{Cartan distribution}~$\mathcal{C}$. This distribution is
integrable in a formal sense, i.e., $[X,Y]\in\mathcal{C}$ for any~$X$,
$Y\in\mathcal{C}$, and its $n$-dimensional integral manifolds are solutions
of~$\mathcal{E}$.

\subsection{Symmetries}
\label{sec:symmetries}

Denote by~$\pi_\infty\colon\mathcal{E}\to M$ the natural projection. A
$\pi_\infty$-vertical vector field~$X$ on~$\mathcal{E}$ is called a
\emph{symmetry} of~$\mathcal{E}$ if it preserves the Cartan distribution,
i.e., if~$[X,\mathcal{C}]\subset\mathcal{C}$. Every symmetry of~$\mathcal{E}$
is of the form
\begin{equation*}
  \re_\phi=\sum D_\sigma(\phi^j)\pd{}{u_\sigma^j},
\end{equation*}
where summation is taken over \emph{internal coordinates} on~$\mathcal{E}$ and
the vector-function~$\phi=(\phi^1,\dots,\phi^m)$ satisfies the equation
\begin{equation*}
  \ell_{\mathcal{E}}(\phi)=0.
\end{equation*}
Here~$\ell_{\mathcal{E}}$ is the \emph{linearization operator} of the vector
function~$F$ restricted to~$\mathcal{E}$. To be more precise, we take the
functions~$F^\alpha$ that define the equation~$\mathcal{E}$ and construct the
matrix $\mathcal{C}$-differential operator
\begin{equation*}
  \ell_F=
  \begin{pmatrix}
    \sum_\sigma\pd{F^1}{u_\sigma^1}D_\sigma&\dots&
    \sum_\sigma\pd{F^1}{u_\sigma^m}D_\sigma\\
    \hdotsfor{3}\\
    \sum_\sigma\pd{F^r}{u_\sigma^1}D_\sigma&\dots
    &\sum_\sigma\pd{F^r}{u_\sigma^m}D_\sigma
  \end{pmatrix}.
\end{equation*}
Since thus defined operator is a $\mathcal{C}$-differential operator, it can
be restricted to~$\mathcal{E}$ and we set
\begin{equation*}
  \ell_{\mathcal{E}}=\eval{\ell_F}_{\mathcal{E}}.
\end{equation*}

The function~$\phi$ is called the \emph{generating function} (or
\emph{section}, or \emph{characteristic}) of the corresponding
symmetry~$\re_\phi$ and we usually make no distinction between symmetries and
their generating functions. The set of symmetries is a Lie algebra
over~$\mathbb{R}$ with respect to the commutator. We denote this algebra
by~$\sym(\mathcal{E})$. The bracket of vector fields induces a bracket of
generating functions by
\begin{equation*}
  \re_{\{\phi_1,\phi_2\}}=[\re_{\phi_1},\re_{\phi_2}],
\end{equation*}
which is called the \emph{Jacobi bracket} and is presented by
\begin{equation*}
  \{\phi_1,\phi_2\}^j=\sum_{\alpha,\sigma}
  \Big(D_\sigma(\phi_1^\alpha)\pd{\phi_2^j}{u_\sigma^\alpha}-
  D_\sigma(\phi_2^\alpha)\pd{\phi_1^j}{u_\sigma^\alpha}\Big)
\end{equation*}
in coordinates.

\subsection{Conservation laws}
\label{sec:conservation-laws}

Consider the space~$\Lambda^1(\mathcal{E})$ of differential $1$-forms
on~$\mathcal{E}$. It consists of finite sums
\begin{equation*}
  \omega=\sum_iA_i\d x^i+\sum_{j,\sigma}B_j^\sigma\d u_\sigma^j,
\end{equation*}
$A_i$ and $B_j^\sigma$ being smooth functions on~$\mathcal{E}$. The
space~$\Lambda^1(\mathcal{E})$ splits naturally into the direct sum
\begin{equation}
  \label{eq:7}
  \Lambda^1(\mathcal{E})=\Lambda_h^1(\mathcal{E})\oplus\Lambda_v^1(\mathcal{E}),
\end{equation}
where
\begin{equation*}
  \Lambda_h^1(\mathcal{E})=\sd{\omega\in\Lambda^1(\mathcal{E})}{\omega=\sum
    A_i\d x^i}
\end{equation*}
is the subspace of \emph{horizontal} forms, while~$\Lambda_v^1(\mathcal{E})$
is generated by the differential forms~$\omega_\sigma^j=\d
u_\sigma^j-\sum_iu_{\sigma i}^j\d x^i$ and is the subspace of \emph{vertical}
(or \emph{Cartan}) forms.

Splitting~\eqref{eq:7} generates the splitting
\begin{equation*}
  \Lambda^s(\mathcal{E})=\sum_{p+q=s}\Lambda_h^q(\mathcal{E})\otimes
  \Lambda_v^p(\mathcal{E}),
\end{equation*}
where
\begin{equation*}
  \Lambda_h^q(\mathcal{E})=
  \underbrace{\Lambda_h^1(\mathcal{E})\wedge\dots\wedge
    \Lambda_h^1(\mathcal{E})}_{q
    \text{ times}},\quad
  \Lambda_v^p(\mathcal{E})=
  \underbrace{\Lambda_v^1(\mathcal{E})\wedge\dots\wedge
    \Lambda_v^1(\mathcal{E})}_{p
    \text{ times}}.
\end{equation*}
Let us introduce the
notation~$\Lambda^{p,q}(\mathcal{E})=\Lambda_h^q(\mathcal{E})\otimes
\Lambda_v^p(\mathcal{E})$.  Consequently, the de~Rham
differential~$\d\colon\Lambda^s(\mathcal{E})\to\Lambda^{s+1}(\mathcal{E})$
splits into the sum of the horizontal
\begin{equation*}
  \d_h\colon\Lambda^{p,q}(\mathcal{E})\to\Lambda^{p,q+1}(\mathcal{E})
\end{equation*}
and vertical
\begin{equation*}
  \d_v\colon\Lambda^{p,q}(\mathcal{E})\to\Lambda^{p+1,q}(\mathcal{E})
\end{equation*}
parts and one has
\begin{equation}
  \label{eq:9}
  [\d_h,\d_v]=0.
\end{equation}
Due to~\eqref{eq:9}, we have a bi-complex structure
on~$\Lambda^*(\mathcal{E})$ which is a particular case of Vinogradov's
$\mathcal{C}$-spectral
sequence~\cite{Vi-DAN-1977,Vi-DAN-1978,Vi-JMAA-1984}. Denote
by~$E_1^{p,q}(\mathcal{E})$ the cohomology of~$\d_h$ at the
term~$\Lambda^{p,q}(\mathcal{E})$. Then~$\d_v$ induces the differentials
\begin{equation*}
  \delta\colon E_1^{p,q}(\mathcal{E})\to E_1^{p+1,q}(\mathcal{E}).
\end{equation*}
The group~$E_1^{0,n-1}(\mathcal{E})$ plays a special role in the theory. Its
elements are called \emph{conservation laws} of~$\mathcal{E}$, while the
group itself is denoted by~$\Cl(\mathcal{E})$. We also shall need the
group~$E_1^{1,n-1}$ whose elements are called \emph{cosymmetries} and which is
denoted by~$\cosym(\mathcal{E})$.

To proceed, we shall need additional constructions. Let~$P$ and~$Q$ be the
spaces of sections of vector bundles
over~$\mathcal{E}$. Let~$\hat{P}=\Hom(P,\Lambda^n(\mathcal{E}))$ and similar
for~$Q$. Then for any $\mathcal{C}$-differential operator~$\Delta\colon P\to
Q$ its \emph{formally adjoint}~$\Delta^*\colon\hat{Q}\to\hat{P}$ is defined by
the \emph{Green formula}
\begin{equation*}
  \langle\Delta^*(\hat{q}),p\rangle-\langle\hat{q},\Delta(p)\rangle=
  \d_h\omega(\hat{q},p), 
\end{equation*}
where~$\omega\colon\hat{Q}\times P\to\Lambda^{n-1}(\mathcal{E})$ is a map
which is a $\mathcal{C}$-differential operator in both arguments
and~$\langle\cdot\,,\cdot\rangle$ denotes the natural pairing. If~$\Delta$ is
given by the matrix~$(\Delta_{ij})$, where~$\Delta_{ij}=\sum_\sigma
a_{ij}^\sigma D_\sigma$, then~$\Delta^*=(\Delta_{ji}^*)$, where
\begin{equation*}
  \Delta_{ji}^*=\sum_\sigma(-1)^{\abs{\sigma}}D_\sigma\circ a_{ji}^\sigma.
\end{equation*}

In what follows, we shall assume that equation at hand satisfies the following
conditions:
\begin{enumerate}
\item The differentials~$\d F^j$ of the functions that define~$\mathcal{E}$
  are linear independent at all points of~$\mathcal{E}$.
\item If~$\Delta$ is a $\mathcal{C}$-differential operator such
  that~$\Delta\circ\ell_{\mathcal{E}}=0$ then~$\Delta=0$.
\item If~$\Delta$ is a $\mathcal{C}$-differential operator such
  that~$\Delta\circ\ell_{\mathcal{E}}^*=0$ then~$\Delta=0$.
\end{enumerate}
If an equation enjoys these conditions then the following statements are
valid\footnote{They follow from Vinogradov's $2$-Line
  Theorem~\cite{Vi-JMAA-1984}.}:
\begin{enumerate}
\item The differential~$\delta\colon\Cl(\mathcal{E})\to\cosym(\mathcal{E})$ is
  monomorphic, i.e.,~$\delta(\omega)=0$ if and only if~$\omega=0$.
\item The group of cosymmetries coincides with the kernel
  of~$\ell_{\mathcal{E}}^*$, i.e.,~$\psi\in\cosym(\mathcal{E})$ if and only
  if~$\ell_{\mathcal{E}}^*(\psi)=0$.
\end{enumerate}

If~$\omega\in\Cl(\mathcal{E})$ is a conservation law then the
cosymmetry~$\delta(\omega)$ is called its \emph{generating function} (or
\emph{generating section}).

\subsection{Differential coverings}
\label{sec:diff-cover}

Let~$\mathcal{E}$ and~$\tilde{\mathcal{E}}$ be two equations. A smooth
map~$\tau\colon\tilde{\mathcal{E}}\to\mathcal{E}$ is called a \emph{morphism}
if it takes the Cartan distribution on~$\tilde{\mathcal{E}}$ to that
on~$\mathcal{E}$. A surjective morphism~$\tau$ is said to be a \emph{covering}
if for any point~$\theta\in\tilde{\mathcal{E}}$ the
differential~$\eval{\d\tau}_\theta$ maps the Cartan
plane~$\mathcal{C}_\theta(\tilde{\mathcal{E}})$
to~$\mathcal{C}_{\tau(\theta)}(\mathcal{E})$ isomorphically. Coordinates along
the fibers of~$\tau$ are called \emph{nonlocal variables} in the covering
under consideration. Let~$\tau'\colon\tilde{\mathcal{E}}'\to\mathcal{E}$ be
another covering. We say that it is \emph{equivalent} to~$\tau$ if there
exists a morphism~$f\colon\tilde{\mathcal{E}}\to\tilde{\mathcal{E}}'$ which is
a diffeomorphism and such that~$\tau=\tau'\circ f$.

If~$D_1,\dots,D_n$ are total derivatives on~$\mathcal{E}$
and~$w^1,\dots,w^r,\dots$ are nonlocal variables then the covering structure
is given by vector fields
\begin{equation}\label{eq:12}
  \tilde{D}_i=D_i+X_i,\qquad i=1,\dots,n,
\end{equation}
where~$X_i=\sum_\alpha X_i^\alpha\partial/\partial w^\alpha$ are
$\tau$-vertical fields that satisfy the condition
\begin{equation}
  \label{eq:10}
  D_i(X_j)-D_j(X_i)+[X_i,X_j]=0,\qquad 1\le i<j\le n.
\end{equation}
A covering is \emph{Abelian} if the coefficients~$X_i^\alpha$ do not depend on
nonlocal variables. In this case,~\eqref{eq:10} amounts to
\begin{equation}
  \label{eq:11}
  D_i(X_j)-D_j(X_i)=0,\qquad 1\le i<j\le n.
\end{equation}
In the particular case of one-dimensional coverings, conditions~\eqref{eq:11}
define a $\d_h$-closed horizontal $1$-form~$\omega_\tau=\sum_iX_i\d x^i$
on~$\mathcal{E}$ and two coverings of this type are equivalent if and only if
the corresponding forms are in the same cohomology class,
i.e.,~$\omega_\tau-\omega_{\tau'}=\d_h(g)$ for some function~$g$. When~$n$
(the number of independent variables) equals two, this establishes a
one-to-one correspondence between the group~$\Cl(\mathcal{E})$ and the
equivalence classes of one-dimensional Abelian coverings over~$\mathcal{E}$.

If~$\tau\colon\tilde{\mathcal{E}}\to\mathcal{E}$ is a covering then symmetries
of~$\tilde{\mathcal{E}}$ are called \emph{nonlocal} $\tau$-symmetries
of~$\mathcal{E}$. Note also that any $\mathcal{C}$-differential
operator~$\Delta$ on~$\mathcal{E}$ can be lifted to a
$\mathcal{C}$-differential operator~$\tilde{\Delta}$
on~$\tilde{\mathcal{E}}$. This is being done by changing total
derivatives~$D_i$ in the local representation of~$\Delta$ to~$\tilde{D}_i$
using~\eqref{eq:12}. In particular, the linearization
operator~$\ell_{\mathcal{E}}$ can be lifted in such a way and solutions of the
equation
\begin{equation*}
  \tilde{\ell}_{\mathcal{E}}(\phi)=0
\end{equation*}
are called (nonlocal) \emph{shadows of symmetries} in the covering~$\tau$. In
a similar way, solutions of
\begin{equation*}
  \tilde{\ell}_{\mathcal{E}}^*(\psi)=0
\end{equation*}
are (nonlocal) \emph{shadows of cosymmetries} in the covering~$\tau$.

\subsection{The $\ell$-covering}
\label{sec:ell-covering}

Let~$\mathcal{E}$ be an equation. Consider a new set of dependent
variables~$q=(q^1,\dots,q^m)$ (in many respects it is convenient to
consider~$q$ as an odd variable), where~$m$ is the number of unknown functions
in~$\mathcal{E}$, and augment the initial equation with
\begin{equation}
  \ell_{\mathcal{E}}(q)=0.\label{eq:14}
\end{equation}
The resulting system, consisting of~$\mathcal{E}$ and equation~\eqref{eq:14},
is called the \emph{$\ell$-covering} of~$\mathcal{E}$. It is an analog of the
tangent bundle for the equation~$\mathcal{E}$. The $\ell$-covering is
important for the subsequent computations due to the following properties.

\subsubsection{Recursion operators for symmetries}
\label{sec:recurs-oper-symm}

Consider a vector-function~$\Phi=(\Phi^1,\dots,\Phi^m)$,
where~$\Phi^j=\sum_{\alpha,\sigma}\Phi_{\alpha,\sigma}^jq_\sigma^\alpha$,
which is a symmetry shadow in the $\ell$-covering. This means that it
satisfies the equation
\begin{equation}\label{eq:17}
  \tilde{\ell}_{\mathcal{E}}(\Phi)=0,
\end{equation}
where~$\tilde{\ell}_{\mathcal{E}}$ is the linearization operator lifted to the
$\ell$-covering. Then it can be shown that the matrix
$\mathcal{C}$-differential
operator~$\mathcal{R}_\Phi=(\sum_\sigma\Phi_{\alpha,\sigma}^jD_\sigma)$ takes
symmetries of~$\mathcal{E}$ to symmetries. In other words,~$\mathcal{R}_\Phi$
is a recursion operator for symmetries.

A recursion operator~$\mathcal{R}$ is called \emph{hereditary} if
\begin{equation*}
  \{\mathcal{R}\phi_1,\mathcal{R}\phi_2\}
  -\mathcal{R}\big(\{\mathcal{R}\phi_1,\phi_2\}+\{\phi_1,\mathcal{R}\phi_2\}
  -\mathcal{R}\{\phi_2,\phi_2\}\big)=0
\end{equation*}
Hereditary operators possess the following property important for
integrability: let~$\phi$ be a symmetry such that
\begin{equation*}
  \re_\phi(\mathcal{R})-[\ell_\phi,\mathcal{R}]=0.
\end{equation*}
Then all symmetries~$\mathcal{R}^i\phi$ pair-wise commute, i.e., form a
commutative hierarchy.

\subsubsection{Symplectic operators}
\label{sec:symplectic-operators}

In a similar way, let us consider now a
vector-function~$\Psi=(\Psi^1,\dots,\Psi^r)$,
where~$\Psi^j=\sum_{\alpha,\sigma}\Psi_{\alpha,\sigma}^jq_\sigma^\alpha$
(recall that~$r$ is the number of the functions~$F^j$ that
define~$\mathcal{E}$), that satisfies the equation
\begin{equation}\label{eq:18}
  \tilde{\ell}_{\mathcal{E}}^*(\Psi)=0,
\end{equation}
where~$\tilde{\ell}_{\mathcal{E}}^*$ is the lift of the
operator~$\ell_{\mathcal{E}}^*$ to the $\ell$-covering. Then the
operator~$\mathcal{S}_\Psi=(\sum_\sigma\Psi_{\alpha,\sigma}^jD_\sigma)$ takes
symmetries of the equation~$\mathcal{E}$ to its cosymmetries.

Let an operator~$\mathcal{S}$ satisfy the condition
\begin{equation}\label{eq:13}
  \mathcal{S}^*\circ\ell_{\mathcal{E}}=\ell_{\mathcal{E}}^*\circ\mathcal{S}.
\end{equation}
Then~$\mathcal{S}$ is identified with a \emph{variational
  $2$-form}~$\Omega_{\mathcal{S}}$ on~$\mathcal{E}$ whose values on symmetries
is given by
\begin{equation*}
  \Omega_{\mathcal{S}}(\phi_1,\phi_2)=\langle\mathcal{S}\phi_1,\phi_2\rangle.
\end{equation*}
This form can be considered as an element of the
group~$E_1^{2,n-1}(\mathcal{E})$ in the term~$E_1$ of the
$\mathcal{C}$-spectral sequence.

Let now~$\omega_1$, $\omega_2$ be two conservation laws such that
\begin{equation*}
  \delta\omega_i=\mathcal{S}\phi_i,\qquad i=1,2,
\end{equation*}
for some~$\phi_1$, $\phi_2\in\sym\mathcal{E}$. Then the bracket
\begin{equation*}
  \{\omega_1,\omega_2\}_{\mathcal{S}}=\Omega_{\mathcal{S}}(\phi_1,\phi_2)
\end{equation*}
is defined. This bracket is skew-symmetric by~\eqref{eq:13} and satisfies the
Jacobi identity if
\begin{equation}
  \label{eq:16}
  \delta\Omega_{\mathcal{S}}=0.
\end{equation}
Operators that enjoy properties~\eqref{eq:13} and~\eqref{eq:16} are called
\emph{symplectic}.

\subsubsection{Nonlocal covectors}
\label{sec:nonlocal-covectors}

Solving equations~\eqref{eq:17} or~\eqref{eq:18} leads often to trivial
results only. The reason is that symplectic and especially recursion operators
are in many cases nonlocal, i.e., contain terms like~$D_x^{-1}$. Such terms
are incorporated into solution by introducing nonlocal variables that amounts
to constructing appropriate coverings. One of the ways to construct the latter
is based on the following fact (see~\cite{KrVe-JGP-2011}): \emph{to any
  cosymmetry of the equation~$\mathcal{E}$ there corresponds a conservation
  law on the $\ell$-covering}. We call these conservation laws \emph{nonlocal
  covectors}. Consequently, if~$n=2$ an Abelian covering corresponds to a
cosymmetry. Numerous
computations~\cite{GoKrVe-AAM-2008,KeKrVe-JGP-2004,KeKrVe-AAM-2006} show that
nonlocal variables arising in such a way are sufficient to find necessary
structures.

\subsection{The $\ell^*$-covering}
\label{sec:ell-covering-1}

Let again~$\mathcal{E}$ be an equation. Consider a another new set of
dependent variables~$p=(p^1,\dots,p^r)$, where~$r$ is the number of
functions~$F^j$ that determine the equation~$\mathcal{E}$, and augment the
initial equation with
\begin{equation}\label{eq:19}
  \ell_{\mathcal{E}}^*(p)=0.
\end{equation}
The resulting system, consisting of~$\mathcal{E}$ and equation~\eqref{eq:19},
is called the \emph{$\ell^*$-covering} of~$\mathcal{E}$. It is an analog of
the cotangent bundle for the equation~$\mathcal{E}$. The $\ell^*$-covering is
also important for the subsequent computations due to the following
properties. Like the variable~$q$ in the $\ell$-covering, it is convenient to
consider the variable~$p$ to be odd.

\subsubsection{Hamiltonian operators}
\label{sec:hamilt-oper}

Consider a vector-function~$\Phi=(\Phi^1,\dots,\Phi^m)$,
where~$\Phi^j=\sum_{\alpha,\sigma}\Phi_{\alpha,\sigma}^jp_\sigma^\alpha$, and
assume that it is a symmetry shadow in the $\ell^*$-covering. This means that
it satisfies the equation
\begin{equation}\label{eq:20}
  \tilde{\ell}_{\mathcal{E}}(\Phi)=0,
\end{equation}
where~$\tilde{\ell}_{\mathcal{E}}$ is the linearization operator lifted to the
$\ell^*$-covering. Then it can be shown that the the matrix
$\mathcal{C}$-differential
operator~$\mathcal{H}_\Phi=(\sum_\sigma\Phi_{\alpha,\sigma}^jD_\sigma)$ takes
cosymmetries of~$\mathcal{E}$ to symmetries.

Solutions to~\eqref{eq:20} of special type are identified with
\emph{variational bivectors}~$\Lambda_{\mathcal{H}}$ on~$\mathcal{E}$. These
solutions must satisfy the condition
\begin{equation*}
  \ell_{\mathcal{E}}\circ\mathcal{H}=\mathcal{H}^*\circ\ell_{\mathcal{E}}^*.
\end{equation*}
In this case, the operation
\begin{equation*}
  \{\omega_1,\omega_2\}_{\mathcal{H}}=
  \langle\mathcal{H}(\delta\omega_1),\delta\omega_2\rangle,\qquad
  \omega_1,\omega_2\in\Cl(\mathcal{E}),
\end{equation*}
defines a skew-symmetric bracket on the space of conservation laws. This
bracket satisfies the Jacobi identity if and only
if~$\ldb\Lambda_{\mathcal{H}},\Lambda_{\mathcal{H}}\rdb=0$,
where~$\ldb\cdot,\cdot\rdb$ is the \emph{variational Schouten bracket}
(see~\cite{KeKrVe-JGP-2004,KrVe-JGP-2011}) on the space of variational
multi-vectors. In this case, one has
\begin{equation*}
  \mathcal{H}\delta(\{\omega_1,\omega_2\}_{\mathcal{H}})=
  \{\mathcal{H}\delta\omega_1,\mathcal{H}\delta\omega_2\}.
\end{equation*}

\subsubsection{Recursion operators for cosymmetries}
\label{sec:recurs-oper-cosymm}

Finally, let us now consider a vector-function~$\Psi=(\Psi^1,\dots,\Psi^r)$,
where~$\Psi^j=\sum_{\alpha,\sigma}\Psi_{\alpha,\sigma}^jp_\sigma^\alpha$, and
assume that it is a cosymmetry shadow in the $\ell^*$-covering. This means
that it satisfies the equation
\begin{equation}\label{eq:22}
  \tilde{\ell}_{\mathcal{E}}^*(\Psi)=0,
\end{equation}
where~$\tilde{\ell}_{\mathcal{E}}$ is the linearization operator lifted to the
$\ell^*$-covering. Then it can be shown that the the matrix
$\mathcal{C}$-differential
operator~$\hat{\mathcal{R}}_\Psi=(\sum_\sigma\Psi_{\alpha,\sigma}^jD_\sigma)$
takes cosymmetries of~$\mathcal{E}$ to cosymmetries. In other
words,~$\hat{\mathcal{R}}$ is a recursion operator for cosymmetries
of~$\mathcal{E}$.

\subsubsection{Nonlocal vectors}
\label{sec:nonlocal-vectors}

Similar to Section~\ref{sec:ell-covering}, equations~\eqref{eq:20}
and~\eqref{eq:22} lead often to trivial results only. The reason is the same:
Hamiltonian and recursion operators are in many cases nonlocal. Such terms are
incorporated into solution by introducing nonlocal variables that amounts to
constructing appropriate coverings. One of the ways to construct the latter is
based on the following fact: \emph{to any symmetry of the
  equation~$\mathcal{E}$ there corresponds a conservation law on the
  $\ell^*$-covering}. We call these conservation laws \emph{nonlocal
  vectors}. Consequently, if~$n=2$ an Abelian covering corresponds to a
symmetry. As
computations~\cite{GoKrVe-AAM-2008,KeKrVe-JGP-2004,KeKrVe-AAM-2006} show,
nonlocal variables arising in such a way are sufficient to find necessary
structures.

\subsection{General computational scheme}
\label{sec:gen-scheme}

In all the computations we did to analyze particular equations
(Equation~\eqref{eq:23} included) we adhered to the following scheme:
\begin{itemize}
\item Extension of the initial equation with a minimal set of nonlocal variables
  (usually associated with conservation laws) to ensure existence of
  nontrivial solutions to the main equations defining integrable structures.
\item Computation of a minimal set of (local and nonlocal) symmetries and
  cosymmetries necessary to (a)~hierarchy generation and (b)~construction of
  nonlocal vectors and covectors.
\item Extension of the $\ell$-covering and and construction of symplectic
  structures and recursion operators for symmetries.
\item Extension of the $\ell^*$-covering and and construction of Hamiltonian
  structures and recursion operators for cosymmetries.
\end{itemize}

\section{Main results}
\label{sec:main-results}

For internal coordinates on~$\mathcal{E}$ we choose the functions
\begin{equation}\label{eq:25}
  u_{k,i}=\pd{^{k+i}u}{x^k\partial y^i},\qquad
  i=0,1,2,\quad k=0,1,\dots,
\end{equation}
and then the total derivatives on~$\mathcal{E}$ acquire the form
\begin{align}\label{eq:26}
  D_x&=\pd{}{x}+\sum_{k\ge0}(u_{k+1,0}\pd{}{u_{k,0}}+u_{k+1,1}\pd{}{u_{k,1}}
  +u_{k+1,2}\pd{}{u_{k,2}}),\\\nonumber
  D_y&=\pd{}{y}+\sum_{k\ge0}(u_{k,1}\pd{}{u_{k,0}}+u_{k,2}\pd{}{u_{k,1}}
  +D_x^k(u_{2,1}^2-u_{3,0}u_{1,2})\pd{}{u_{k,2}}).
\end{align}

Equation~\eqref{eq:23} is homogeneous with respect to the following
weights
\begin{equation*}
  \abs{u}=0,\quad\abs{x}=-1,\quad\abs{y}=-4.
\end{equation*}

\subsection{Conservation laws and Abelian coverings}
\label{sec:ch-appl:--cl-2-ass}

In the sequel, we shall need nonlocal variables that will be denoted
by~$Q_{i,j}$. The second subscript here indicates the weight of the
variable, while first the one corresponds to the \emph{level of
  nonlocality}. By the latter we mean the following. The variables of zero
level are determined by local functions on~$\mathcal{E}$:
\begin{gather*}
    \pd{Q_{0,7}}{x} =-u_{0,1}u_{4,0} + u_{0,2},\quad
    \pd{Q_{0,7}}{y} =-u_{0,1}u_{3,1} - u_{0,2}u_{3,0} + u_{1,1}u_{2,1},\\
    \pd{Q_{0,9}}{x} =2u_{0,2}u_{2,0} + u_{1,1}^2 - u_{2,0}^2u_{2,1},\quad
    \pd{Q_{0,9}}{y} =2u_{0,2}u_{1,1} - u_{1,2}u_{2,0}^2,
    \intertext{and}
    \pd{Q_{0,12}}{x} =u_{1,1} (u_{0,2} - u_{2,0}u_{2,1}),\quad
    \pd{Q_{0,12}}{y} = \frac{1}{2}(u_{0,2}^2 - 2, u_{1,1}, u_{1,2}u_{2,0}).
\end{gather*}
The variables of level~$1$ are determined by local functions and by variables
of zero level:
\begin{gather*} 
    \pd{Q_{1,3}}{x} =2u_{0,1} - u_{2,0}^2,  \quad 
    \pd{Q_{1,3}}{y} =2(Q_{0,7} + u_{0,1}u_{3,0} - u_{1,1}u_{2,0}),\\
    \pd{Q_{1,6}}{x}  = Q_{0,7} + u_{0,1}u_{3,0}, \quad  
    \pd{Q_{1,6}}{y}  = \frac{1}{2}u_{1,1}^2,
\end{gather*}
and
\begin{align*}
  \pd{Q_{1,8}}{x} &=Q_{0,9} - 2u_{0,1}u_{1,0}u_{4,0} -
  4u_{0,1}u_{20}u_{30} + 2u_{0,2}u_{1,0}, \\
  \pd{Q_{1,8}}{y} &= 4Q_{0,12}-2u_{0,1}u_{1,0}u_{3,1}
  -2u_{0,1}u_{1,1}u_{3,0}-2u_{0,1}u_{2,0}u_{2,1} \\
  &-2u_{0,2}u_{1,0}u_{3,0}-u_{0,2}u_{2,0}^2+2u_{1,0}u_{1,1}u_{2,1} +
  2u_{1,1}^2u_{2,0}.
\end{align*}
There exist deeper level nonlocalities, such as
\begin{align*}
  \pd{Q_{2,5}}{x}&=-18Q_{1,6}-2u_{0,1}u_{2,0}+4u_{1,0}u_{1,1}+u_{2,0}^3, \\
  \pd{Q_{2,5}}{y}&=-3Q_{0,9}-2u_{0,1}u_{1,1}+4u_{0,2}u_{1,0}
\end{align*}
and
\begin{align*}
  \pd{Q_{2,7}}{x}&=-40Q_{0,7}u_{1,0}-10Q_{1,8}-10u_{0,1}^2
  -60u_{0,1}u_{1,0}u_{3,0}\\
  &+u_{1,0}u_{2,0}^2u_{3,0}-\frac{1}{2}u_{20}^4, \\
  \pd{Q_{2,7}}{y}&=-40Q_{0,7}u_{01}-10Q_{1,11}-30u_{0,1}^2u_{3,0}
  +20u_{0,1}u_{1,1}u_{2,0}\\
  &+10u_{0,1}u_{2,0}^2u_{3,0}-30u_{1,0}u_{1,1}^2+u_{1,0}u_{2,0}^2u_{2,1}-3u_{1,1}u_{2,0}^3,
\end{align*}
as well as
\begin{align*}
  \label{eq:ch-appl:269}
    \pd{Q_{3,4}}{x}&=\frac{1}{3}Q_{2,5}+Q_{1,3}u_{2,0}
    -\frac{4}{3}u_{0,1}u_{1,0},\\
    \pd{Q_{3,4}}{y}&=2Q_{0,7}u_{1,0}+Q_{1,3}u_{1,1}-Q_{1,8}
    -2u_{0,1}^2-u_{0,1}u_{2,0}^2.
\end{align*}

\begin{remark}
  \label{exrc:ch-appl:66}
  The zero level nonlocal variables are associated to conservation laws of
  equation~\eqref{eq:23}. For example, to~$Q_{0,7}$ there
  corresponds the conservation law
  \begin{equation*}
    \omega_{0,7}=(-u_{0,1}u_{4,0} + u_{0,2})\,dx
    +(-u_{0,1}u_{3,1} - u_{0,2}u_{3,0} + u_{1,1}u_{2,1})\,dy.
  \end{equation*}
  The first level nonlocal variables are associated to conservation laws of
  the Abelian coverings determined by the zero level variables, etc.
\end{remark}

\begin{remark}
  Of course, the list of nonlocal variables above is not exhaustive at all. We
  described only those ones that are used to construct the necessary nonlocal
  symmetries (\S~\ref{sec:ch-appl:sym-2-ass}) and cosymmetries
  (\S~\ref{sec:ch-appl:cosym-2-ass}); see also \S~\ref{sec:gen-scheme}. New
  nonlocalities do arise under the actions of recursion operators, but there
  is no need to describe them explicitly here.
\end{remark}

\subsection{Symmetries}
\label{sec:ch-appl:sym-2-ass}

This direct computation is needed by the two reasons: to construct nonlocal
vectors (see Subsection~\ref{sec:nonlocal-vectors}) and to use the obtained
symmetries as `seeds' for the hierarchies.

The linearization of equation~\eqref{eq:23} has the form
\begin{equation}\label{eq:27}
  D_y^3(\phi)-2u_{2,1}D_x^2D_y(\phi)+u_{1,2}D_x^3(\phi)+u_{3,0}D_xD_y^2(\phi) =0,
\end{equation}
where the total derivatives~$D_x$ and~$D_y$ are given by~\eqref{eq:26}.

Solving~\eqref{eq:27}, we get the following solutions\footnote{In the notation
  for symmetries, the superscript indicates the polynomial order of a symmetry
  with respect to~$x$ and~$y$, the first subscript equals the weight, while
  the second one, if any, is the number of a symmetry in the set given by
  particular weight and order.}.

\subsubsection{Symmetries of degree~$0$\textup{:}}
\label{sec:ch-appl:--0}

\begin{gather*}
  \phi_0^0  = 1,\quad
  \phi_1^0  = u_{1,0},\quad
  \phi_4^0 = u_{0,1},\quad
  \phi_5^0  = Q_{2,5} + 8u_{0,1}u_{1,0},\\
  \phi_8^0  = -2Q_{0,7}u_{1,0}+Q_{1,8}-2u_{0,1}^2+u_{0,1}u_{2,0}^2.
\end{gather*}

\subsubsection{Symmetries of degree~$1$\textup{:}}
\label{sec:ch-appl:--1}

\begin{gather*}
  \phi_{-4}^1   = y,\quad
  \phi_{-1}^1   = x,\quad  \\
  \phi_{0,1}^1  =x u_{1,0} - 4 u,\qquad \phi_{0,2}^1  =y u_{0,1} + u,\\
  \phi_3^1     = 4xu_{0,1} - Q_{1,3},\\
  \phi_4^1  = x(Q_{2,5}+8u_{0,1}u_{1,0})-3(8Q_{3,4}-3Q_{1,3}u_{1,0}+16uu_{0,1}).
\end{gather*}

\subsubsection{Symmetries of degree~$2$\textup{:}}
\label{sec:ch-appl:--2}

\begin{gather*}
  \phi_{-4}^{2}   = y,\quad
  \phi_{-8}^{2}  = y^2,\quad
  \phi_{-5}^{2}  = x y,\quad
  \phi_{-2}^{2}  = x^2,\\
  \phi_{-1}^{2}  = x^2 u_{1,0} + 4 x y u_{0,1} - 4 x u - y Q_{1,3},\\
  \phi_{2}^{2}  = 2 x^2 u_{0,1} - x Q_{1,3} - u_{1,0}^2.
\end{gather*}

\subsubsection{Symmetries of degree~$3$\textup{:}}
\label{sec:ch-appl:--3}

\begin{gather*}
  \phi^{3}_{-3}  = x^3 - 2 y u_{1,0},\\
  \phi^{3}_1 = 12x^3u_{0,1}-9x^2Q_{1,3}-18xu_{1,0}^2 - 2 y (Q_{2,5}+8 u_{0,1}
  u_{1,0}) + 24uu_{1,0}.
\end{gather*}

\subsubsection{Symmetries of degree~$4$\textup{:}}
\label{sec:ch-appl:--4}

We shall also need one symmetry of order~$4$, which is of the form
\begin{equation*}
  \phi_{-4}^4=x^4 - 8 x y u_{1,0} - 8 y^2 u_{0,1} + 16 y u.
\end{equation*}

\subsection{Cosymmetries}
\label{sec:ch-appl:cosym-2-ass}

The reasons to compute cosymmetries explicitly are similar to those
indicated in Subsection~\ref{sec:ch-appl:sym-2-ass}.

To find cosymmetries, we are to solve the equation adjoint
to~\eqref{eq:27}, i.e.,
\begin{equation}\label{eq:28}
  D_y^3(\psi)-2D_x^2D_y(u_{2,1}\psi)+D_x^3(u_{1,2}\psi)+D_xD_y^2(u_{3,0}\psi)=0.
\end{equation}
Using the notation similar to the one from
Section~\ref{sec:ch-appl:sym-2-ass}, let us some computational results needed
below.

\subsubsection{Cosymmetries of degree~$0$\textup{:}}
\label{sec:ch-appl:--0-1}

\begin{gather*}
  \psi_0^0  = 1,\quad
  \psi_2^0  = u_{2,0},\quad
  \psi^0_5  = u_{1,1},\\
  \psi_6^0  = -18Q_{1,6}+6u_{0,1}u_{2,0}+12u_{1,0}u_{1,1}+u_{2,0}^3,\\
  \psi_9^0  = 2Q_{0,7}u_{2,0}-Q_{0,9}+4u_{0,1}u_{1,1}+2u_{0,1}u_{2,0}u_{3,0}
  -u_{1,1}u_{2,0}^2.
\end{gather*}

\subsubsection{Cosymmetries of degree~$1$\textup{:}}
\label{sec:ch-appl:--1-1}

\begin{gather*}
  \psi_{-4}^1  = y,\quad
  \psi_{-1}^1  = x,\quad
  \psi_{1,1}^1  = xu_{2,0} - 3 u_{1,0},\\
  \psi_{1,2}^1  = y u_{1,1} + u_{1,0},\quad
  \psi_{4}^1  = 4 x u_{1,1} + 2 u_{0,1} + u_{2,0}^2.
\end{gather*}

\subsubsection{Cosymmetries of degree~$2$\textup{:}}
\label{sec:ch-appl:--2-1}

\begin{gather*}
  \psi_{-2}^2  = 3 x^2 - 2 y u_{2,0},\\
  \psi_0^2  = x^2u_{2,0}+4xyu_{1,1}-2xu_{1,0}+y(2u_{0,1}+u_{2,0}^2)-4 u,\\
  \psi_3^2  = 2x^2u_{1,1}+x(2u_{0,1}+u_{2,0}^2)-Q_{1,3}-2u_{1,0}u_{2,0}.
\end{gather*}

\subsubsection{Cosymmetries of degree~$3$\textup{:}}
\label{sec:ch-appl:--3-1}

\begin{gather*}
  \psi_{-3}^3  = 2u_{1,0}y-2u_{1,1}y^2-2u_{2,0}xy+x^3,\\
  \psi_2^3  = x^3 u_{1,1}+x^2\left(\frac{3}{2}u_{0,1}+\frac{3}{4}u_{2,0}^2\right)
  -x\left(\frac{3}{2}Q_{1,3}+3u_{1,0}u_{2,0}\right)\\
  + y\left(3Q_{1,6}-u_{0,1}u_{2,0}-2u_{1,0}u_{1,1}-\frac{1}{6}u_{2,0}^3\right)
  + (2uu_{2,0}+\frac{1}{2}u_{1,0}^2).
\end{gather*}

\subsection{The $\ell$-covering}
\label{sec:ch-appl:ell-2-ass}

This covering is determined by the system of equations
\begin{align*}
    &u_{yyy}-u_{xxy}^2 +u_{xxx} u_{xyy}=0,\\
    &q_{yyy}-2u_{xxy}q_{xxy}+u_{xyy}q_{xxx} +u_{xxx}q_{xyy}=0,
\end{align*}
where~$q$ is an odd variable along the fiber of the covering.

Internal coordinates on the space of the $\ell$-covering are
functions~\eqref{eq:25} together with
\begin{equation}
  \label{eq:ch-appl:281}
  q_{k,i}=\pd{^{k+i}q}{x^k\partial y^i},\qquad i=0,1,2,\quad
  k=0,1,\dots,
\end{equation}
while the total derivatives in these coordinates take the form
\begin{align*}
  D_x&=\pd{}{x}+\sum_{k\ge0}\Big(u_{k+1,0}\pd{}{u_{k,0}}+u_{k+1,1}\pd{}{u_{k,1}}
  +u_{k+1,2}\pd{}{u_{k,2}}\\
  &+q_{k+1,0}\pd{}{q_{k,0}}+q_{k+1,1}\pd{}{q_{k,1}}
  +q_{k+1,2}\pd{}{q_{k,2}}\Big),\\
  D_y&=\pd{}{y}+\sum_{k\ge0}\Big(u_{k,1}\pd{}{u_{k,0}}+u_{k,2}\pd{}{u_{k,1}}
  +D_x^k(u_{2,1}^2-u_{3,0}u_{1,2})\pd{}{u_{k,2}}\\
  &+q_{k,1}\pd{}{q_{k,0}}+q_{k,2}\pd{}{q_{k,1}}
  +D_x^k(2u_{2,1}q_{2,1}-u_{1,2}q_{3,0}-u_{3,0}q_{1,2})\pd{}{q_{k,2}}\Big).
\end{align*}

By the general theory~\cite{KrVe-JGP-2011}, to any cosymmetry~$\psi$ of the
initial equation there corresponds a conservation law on the $\ell$-covering
(a nonlocal form). Denote by~$\bom_\psi$ the corresponding nonlocal
variable. In the case of Equation~\eqref{eq:23}, this variable is defined by
the relations
\begin{align}\label{eq:30}\nonumber
    \pd{\bom_\psi}{x}&=\psi q_{0,2} + a_{0,1} q_{0,1} + a_{0,0}q,\\
    \pd{\bom_\psi}{y}&=b_{0,2}q_{0,2}+b_{1,1}q_{1,1}+b_{2,0}q_{2,0}+b_{0,1}q_{0,1}
    +b_{1,0}q_{1,0}+b_{0,0}q,
\end{align}
where
\begin{gather}\label{eq:29}
  b_{0,2}=-u_{3,0}\psi,\quad b_{1,1}=2u_{2,1}\psi,\quad
  b_{2,0}=-u_{1,2}\psi,\\\nonumber b_{0,1}=-D_x(b_{1,1}),\quad
  b_{1,0}=-D_x(b_{2,0}),\\
  \nonumber b_{0,0}=-D_x(b_{1,0})
\end{gather}
and
\begin{equation}\label{eq:31}
  a_{0,1}=D_x(b_{0,2}) - D_y(\psi),\quad
  a_{0,0}=D_x(b_{0,1}) - D_y(a_{0,1}).
\end{equation}

Below we use the notation~$\bom_{i,j}^k=\bom_{\psi_{i,j}^k}$

\subsubsection{Recursion operators for symmetries}
\label{sec:ch-appl:--rec-sym-2-ass}

To find recursion operators for symmetries of Equation~\eqref{eq:23},
we solve the equation
\begin{equation*}
  \tilde{\ell}_{\mathcal{E}}(\Phi)=0,
\end{equation*}
where~$\tilde{\ell}_{\mathcal{E}}$ is the linearization operator~\eqref{eq:27}
in the $\ell$-covering extended by nonlocal forms and~$\Phi$ is a function on
this extension.

The simplest nontrivial solution is of the form
\begin{align}\label{eq:34}
  \nonumber
  \Phi_1&= \bom_{-3}^3-x\bom_{-2}^2+4y\bom_{1,2}^1+2y\bom_{1,1}^1+3x^2\bom_{-1}^1\\
  &-2u_{1,0}\bom_{-4}^1-2y^2\bom_5^0-2xy\bom_2^0+(2u_{1,0}y-x^3)\bom_0^0.
\end{align}

Using the first of Equations~\eqref{eq:30}, we put into
correspondence to every nonlocal form~$\bom_\psi$ the operator
\begin{equation}\label{eq:32}
  \mathcal{D}_\psi=D_x^{-1}\circ(\psi D_y^2+a_{0,1}D_y+a_{0,0})
\end{equation}
where the coefficients are determined using relations~\eqref{eq:29}
and~\eqref{eq:31},~i.e.,
\begin{gather*}
  a_{0,1}=-D_x(u_{3,0}\psi)-D_y(\psi),\\
  a_{0,0}=-2D_x^2(u_{2,1}\psi)+D_xD_y(u_{3,0}\psi)+D_y^2(\psi).
\end{gather*}
Then the recursion operator
\begin{align}\label{eq:33}
  \nonumber
  \mathcal{R}_1&= \mathcal{D}_{\psi_{-3}^3}-x\mathcal{D}_{\psi_{-2}^2}
  +4y\mathcal{D}_{\psi_{1,2}^1}+2y\mathcal{D}_{\psi_{1,1}^1}
  +3x^2\mathcal{D}_{\psi_{-1}^1}\\
  &-2u_{1,0}\mathcal{D}_{\psi_{-4}^1}-2y^2\mathcal{D}_{\psi_5^0}
  -2xy\mathcal{D}_{\psi_2^0}+(2u_{1,0}y-x^3)\mathcal{D}_{\psi_0^0}
\end{align}
corresponds to solution~\eqref{eq:34}.

\begin{remark}
  \label{rem:ch-appl:23}
  Since the variables~$x$ and~$y$ in Equation~\eqref{eq:23} ``enjoy
  equal rights'', one can put into correspondence to the nonlocal
  form~$\bom_\psi$, using the second equality in~\eqref{eq:30}, the
  operator
  \begin{equation*}
    \mathcal{D}'_\psi=D_x^{-1}\circ(b_{0,2}D_y^2+b_{1,1}D_xD_y
    +b_{2,0}D_x^2+b_{0,1}D_y+b_{1,0}D_x+b_{0,0})
  \end{equation*}
  and construct a recursion operator~$\mathcal{R}'_1$ similar to
  operator~\eqref{eq:33}.

  The action of operators~$\mathcal{R}_1$ and~$\mathcal{R}'_1$ on symmetries
  of Equation~\eqref{eq:23} is the same.
\end{remark}

\subsubsection{Symplectic structures}
\label{sec:ch-appl:-symp-2-ass}

To find symplectic structures, we solve the equation
\begin{equation*}
  \tilde{\ell}_{\mathcal{E}}^*(\Psi)=0,
\end{equation*}
where, similar to Section~\ref{sec:ch-appl:--rec-sym-2-ass},
$\tilde{\ell}_{\mathcal{E}}^*$ is the operator~\eqref{eq:28} on the
$\ell$-covering extended by the nonlocal forms, while~$\Psi$ is a function on
this extension.

Below we present the first two solutions of this equation. The simplest one is
\begin{equation*}
  \Psi_1=\bom_{1,0},
\end{equation*}
and the corresponding symplectic structure
\begin{equation}\label{eq:35}
  \mathcal{S}_1=D_x.
\end{equation}
The next solution is nonlocal:
\begin{equation*}
  \Psi_2=\bom_{-1}^2-6x\bom_{-1}^1+2u_{2,0}\bom_{-4}^1+2y\bom_2^0
  -(2u_{2,0}y-3 x^2)\bom_0^0.
\end{equation*}
The corresponding symplectic operator~$\mathcal{S}_2\colon\sym\mathcal{E}
\to\cosym\mathcal{E}$ is
\begin{equation*}
  \mathcal{S}_2=
  \mathcal{D}_{\psi_{-1}^2}-6x\mathcal{D}_{\psi_{-1}^1}
  +2u_{2,0}\mathcal{D}_{\psi_{-4}^1}
  +2y\mathcal{D}_{\psi_2^0}-(2u_{2,0}y-3 x^2)\mathcal{D}_{\psi_0^0},
\end{equation*}
where the operators~$\mathcal{D}_\psi$ are defined by Equation~\eqref{eq:32}.

\subsection{The $\ell^*$-covering}
\label{sec:ch-appl:ellstar-cov-2-ass}

As it was indicated above, this covering is obtained by adding to the initial
equation another one, which is adjoint to the linearization. In other words,
this covering is described by the equations
\begin{align*}\nonumber
  &u_{yyy}-u_{xxy}^2 +u_{xxx} u_{xyy}=0,\\
  &u_{xxyy}p_{xx}-2u_{xxxy}p_{xy}+u_{xxxx}p_{yy}\\
  &\qquad+u_{xyy}p_{xxx}-2u_{xxy}p_{xxy}+u_{xxx}p_{xyy}+p_{yyy}=0,
\end{align*}
where~$p$ is a new odd variable.

Internal coordinates in the space of the $\ell^*$-covering are
functions~\eqref{eq:25} together with the functions
\begin{equation*}
  p_{k,i}=\pd{^{k+i}p}{x^k\partial y^i},\qquad
  i=0,1,2,\quad k=0,1,\dots,
\end{equation*}
while the total derivatives in these coordinates are of the form
\begin{align*}
  D_x&=\pd{}{x}+\sum_{k\ge0}\Big(u_{k+1,0}\pd{}{u_{k,0}}+u_{k+1,1}\pd{}{u_{k,1}}
  +u_{k+1,2}\pd{}{u_{k,2}}\\
  &+p_{k+1,0}\pd{}{p_{k,0}}+p_{k+1,1}\pd{}{p_{k,1}}
  +p_{k+1,2}\pd{}{p_{k,2}}\Big),\\
  D_y&=\pd{}{y}+\sum_{k\ge0}\Big(u_{k,1}\pd{}{u_{k,0}}+u_{k,2}\pd{}{u_{k,1}}
  +D_x^k(u_{2,1}^2-u_{3,0}u_{1,2})\pd{}{u_{k,2}}\\
  &+p_{k,1}\pd{}{p_{k,0}}+p_{k,2}\pd{}{p_{k,1}}
  -D_x^k(u_{2,2}p_{2,0}-2u_{3,1}p_{1,1}+u_{4,0}p_{0,2}\\ &
  +u_{1,2}p_{3,0}-2u_{2,1}p_{2,1}+u_{3,0}p_{1,2})\pd{}{p_{k,2}}\Big).
\end{align*}

Let~$\phi$ be a symmetry of Equation~\eqref{eq:23}. Then
(see~\cite{KrVe-JGP-2011}) a conservation law on the $\ell^*$-covering
corresponds to this symmetry and, consequently a nonlocal variable, which we
denote by~$\bop_\phi$ and call a \emph{nonlocal vector}. For
Equation~\eqref{eq:23}, the correspondence~$\phi\mapsto\bop_\phi$ is given by
the relations
\begin{align}\label{eq:36}
  \nonumber
  \pd{\bop_\phi}{x}&=\phi p_{0,2}+a_{0,1}p_{0,1}+a_{0,0}p,\\
  \pd{\bop_\phi}{y}&=b_{0,2}p_{0,2}+b_{1,1}p_{1,1}+b_{2,0}p_{2,0}+b_{0,1}p_{0,1}+
  b_{1,0}p_{1,0}+b_{0,0}p,
\end{align}
where
\begin{gather}
  \nonumber b_{0,2} = -u_{3,0}\phi,\quad b_{1,1} = 2u_{2,1}\phi,\quad b_{2,0}
  = -u_{1,2}\phi,\\\nonumber b_{0,1} = -D_x(b_{1,1})+2u_{3,1}\phi,\quad
  b_{1,0} = -D_x(b_{2,0})-u_{2,2}\phi,\\\label{eq:1}
  b_{0,0} = -D_x(b_{1,0})
\end{gather}
and
\begin{equation*}
  a_{0,1} = D_x(b_{0,2})-D_y(\phi)+u_{4,0}\phi,\quad
  a_{0,0} = D_x(b_{0,1})-D_y(a_{0,1}).
\end{equation*}

We use the notation~$\bop_{i,j}^k=\bop_{\phi_{i,j}^k}$ below.

To describe the subsequent results, let us put into correspondence to a
nonlocal vector~$\bop_\phi$ the operator
\begin{equation}\label{eq:37}
  \mathcal{D}_\phi=D_x^{-1}\circ(\phi D_y^2+a_{0,1}D_y+a_{0,0}),
\end{equation}
see the first of Equations~\eqref{eq:36}. Here, due to
relations~\eqref{eq:1} and~\eqref{eq:37}, the
coefficients~$a_{0,0}$ and~$a_{0,1}$ are of the form
\begin{gather*}
  a_{0,0}=-2u_{2,1}D_x^2(\phi)+u_{3,0}D_xD_y(\phi)+D_y^2(\phi)-u_{3,1}D_x(\phi),\\
  a_{0,1}=-u_{3,0}D_x(\phi)-D_y(\phi).
\end{gather*}

\subsubsection{Hamiltonian structures}
\label{sec:ch-appl:-Ham-2-ass}

Similar to Sections~\ref{sec:ch-appl:--rec-sym-2-ass}
and~\ref{sec:ch-appl:-symp-2-ass}, Hamiltonian structures are solutions of the
equation
\begin{equation}\label{eq:38}
  \tilde{\ell}_{\mathcal{E}}(\Phi)=0,
\end{equation}
where the operator~$\tilde{\ell}_{\mathcal{E}}$ is the linearization lifted to
the $\tilde{\ell}^*$-covering extended by nonlocal vectors.

The simplest solution of Equation~\eqref{eq:38} is
\begin{equation*}
  \Phi_0=\bop_{-8}^2 - 2y\bop_{-4}^1 + y^2\bop_0^0,
\end{equation*}
to which the operator~$\mathcal{H}_0\colon\cosym\mathcal{E}\to\sym\mathcal{E}$ 
\begin{equation*}
  \mathcal{H}_0=\mathcal{D}_{\phi_{-8}^2}-2y\mathcal{D}_{\phi_{-4}^1}+
  y^2\mathcal{D}_{\phi_0^0}
\end{equation*}
corresponds. The next solution is much more complicated and has the form
\begin{align*}
  \Phi_1&=\bop_{-4}^4 -4x\bop_{-3}^3 +6x^2\bop_{-2}^2-8u_{1,0}\bop_{-5}^2
  -8u_{0,1}\bop_{-8}^2+16y\bop_{0,2}^1\\
  &+(8xu_{1,0}+16yu_{0,1}-16u)\bop_{-4}^1+8y\bop_{0,1}^1-(4x^3-8yu_{1,0})\bop_{-1}^1\\
  &-8y^2\bop_4^0-8xy\bop_1^0+(x^4-8xyu_{1,0}-8y^2u_{0,1}+16yu)\bop_0^0
\end{align*}
and the operator
\begin{align*}
  \mathcal{H}_1&=\mathcal{D}_{\phi_{-4}^4}-4x\mathcal{D}_{\phi_{-3}^3}
  +6x^2\mathcal{D}_{\phi_{-2}^2}-8u_{1,0}\mathcal{D}_{\phi_{-5}^2}
  -8u_{0,1}\mathcal{D}_{\phi_{-8}}^2+16y\mathcal{D}_{\phi_{0,2}^1}\\
  &+(8xu_{1,0}+16yu_{0,1}-16u)\mathcal{D}_{\phi_{-4}^1}
  +8y\mathcal{D}_{\phi_{0,1}^1}-(4x^3-8yu_{1,0})\mathcal{D}_{\phi_{-1}^1}\\
  &-8y^2\mathcal{D}_{\phi_4^0}-
  8xy\mathcal{D}_{\phi_1^0}+(x^4-8xyu_{1,0}-8y^2u_{0,1}+16yu)\mathcal{D}_{\phi_0^0}.
\end{align*}
corresponds to this solution.  Here and below the operators~$\mathcal{D}_\phi$
are defined by Equality~\eqref{eq:37}.

\begin{remark}
  \label{rem:ch-appl:24}
  The form of the above presented solutions is determined by the choice of
  nonlocal vectors in the
  $\ell^*$-covering. They, in turn, are due to a basis in the space of
  symmetries. Actually, Equation~\eqref{eq:38} has a simpler
  solution~$\Phi'_0=\bop'_1$, where the nonlocal variable~$\bop'_1$ is defined
  by the system
  \begin{align*}
    &\pd{\bop'_1}{x}=p,&&\pd{\bop'_1}{y}=\bop'_2,\\
    &\pd{\bop'_2}{x}=p_{0,1},&&\pd{\bop'_2}{y}=\bop'_3,\\
    &\pd{\bop_3}{x}=p_{0,2},&&\pd{\bop'_3}{y}=
    -u_{3,0}p_{0,2}+2u_{2,1}p_{1,1}-u_{1,2}p_{2,0}.
  \end{align*}
  To this solution there corresponds the Hamiltonian operator
  \begin{equation*}
    \mathcal{H}'_0=D_x^{-1},
  \end{equation*}
  which is the inverse to the symplectic operator~\eqref{eq:35}. This operator
  corresponds to the Hamiltonian operator~$J_0$
  from~\cite{KaNu-JPhA-1998}. The operator~$\mathcal{H}_1$ that explicitly
  depends on~$x$ and~$y$ seems to be new.
\end{remark}

\begin{remark}
  \label{exrc:ch-appl:72}
  There is a relation
  \begin{equation*}
    \mathcal{H}_1=\mathcal{R}_1\circ\mathcal{H}_0,
  \end{equation*}
  between the two Hamiltonian structures, where~$\mathcal{R}_1$ is the
  recursion operator given by Equality~\eqref{eq:33}.
\end{remark}

\subsubsection{Recursion operators for cosymmetries}
\label{sec:ch-appl:--rec-cosym-2-ass}

To conclude our study, we consider finally the equation
\begin{equation*}
  \tilde{\ell}_{\mathcal{E}}^*(\Psi)=0
\end{equation*}
on the $\ell^*$-covering extended by nonlocal vectors. Here is its simplest
nontrivial solution
\begin{align*}
  \Psi_0&=\bop_{-3}^3-3x\bop_{-2}^2+2u_{2,0}\bop_{-5}^2+2u_{1,1}\bop_{-8}^2\\
  &+2(u_{1,0}+2u_{1,1}y-u_{2,0}x)\bop_{-4}^1-(2u_{2,0} y-3x^2)\bop_{-1}^1+2y\bop_1^0\\
  &-(2u_{1,0}y-2u_{1,1}y^2-2u_{2,0}xy+x^3)\bop_0^0.
\end{align*}
To this solution there corresponds the recursion
operator~$\hat{R}_0\colon\cosym\mathcal{E}\to\cosym\mathcal{E}$ of the form
\begin{align*}
  \hat{R}_0&=\mathcal{D}_{\phi_{-3}^3}-3x\mathcal{D}_{\phi_{-2}^2}
  +2u_{2,0}\mathcal{D}_{\phi_{-5}}^2+2u_{1,1}\mathcal{D}_{\phi_{-8}^2}\\
  &+2(u_{1,0}-2u_{1,1}y-u_{2,0}x)\mathcal{D}_{\phi_{-4}^1}
  -(2u_{2,0} y-3x^2)\mathcal{D}_{\phi_{-1}^1}+2y\mathcal{D}_{\phi_1^0}\\
  &-(2u_{1,0}y-2u_{1,1}y^2-2u_{2,0}xy+x^3)\mathcal{D}_{\phi_0^0}.
\end{align*}

\begin{remark}
  In the case of evolution equations, existence of a commutative hierarchy
  means that the initial equation possesses ``higher analogs'' that are
  obtained by the action of a recursion operator. In the general case, there
  exists no well defined action of recursion operators on the equation and
  thus such a construction is impossible. Nevertheless, one can consider the
  following alternative scheme: (1)~to pass, if possible, to the evolutionary
  presentation of the equation at hand, (2)~to construct ``higher analogs'' in
  this presentation and (3)~return back to the initial variables. We did not
  set the question whether such a scheme is invariant (i.e., whether the
  result is completely determined by the initial equation or depends on the
  procedure). Probably, an answer to this question will lead to a deeper
  understanding of integrability for general equations.
\end{remark}

\end{document}